\documentclass[11pt]{article}
\usepackage{a4wide}
\usepackage[svgnames]{xcolor}
\usepackage{enumerate}
\usepackage{amsmath,amssymb,amsthm}
\usepackage{graphicx}
\usepackage{float}
\usepackage{braket,bm}
\usepackage{url}
\usepackage{mathtools}
\usepackage{multirow}
\usepackage[bookmarks=true,bookmarksnumbered=true,setpagesize=false]{hyperref}

\usepackage{fancybox}
\usepackage{comment}

\usepackage{fontawesome}
\newcommand{\fs}{\text{\faStarO}}

\newcommand{\ub}{\underbrace}

\newcommand{\bp}{\begin{pmatrix}}
\newcommand{\ep}{\end{pmatrix}}
\newcommand{\bb}{\begin{bmatrix}}
\newcommand{\eb}{\end{bmatrix}}
\newcommand{\bmat}[1]{\begin{bmatrix}#1\end{bmatrix}}

\newcommand{\df}{\text{d}}

\newcommand{\bs}[1]{\boldsymbol{#1}}

\newcommand{\al}[1]{\begin{align}#1\end{align}}

\newcommand{\ab}[1]{\left\lvert#1\right\rvert}

\newcommand{\paren}[1]{\left (#1\right) }
\newcommand{\pn}{\paren}
\newcommand{\sqbr}[1]{\left[#1\right]}
\newcommand{\fn}[1]{\!\paren{#1}} % function
\newcommand{\fnl}[1]{\!\sqbr{#1}} % functional

\newcommand{\Ab}[1]{\bigl\lvert#1\bigr\rvert}

\newcommand{\Paren}[1]{\bigl (#1\bigr) }
\newcommand{\Pn}{\Paren}

\usepackage{mathrsfs}
\newcommand{\mc}{\mathcal}
\newcommand{\ms}{\mathscr}

\newcommand{\nn}{\nonumber\\}

\newcommand{\p}{\partial}

\newcommand{\h}{\hat}

\newcommand{\anticommutator}[2]{\left\{#1,\,#2\right\}}

\usepackage{fancybox}

\usepackage{amsthm}
\theoremstyle{definition}

\newcommand{\ov}{\over}
\newcommand{\tx}{\text}

\newcommand{\wh}{\widehat}

\newcommand{\D}{\text{D}}
\newcommand{\s}{\text{S}}

\newcommand{\ePbar}{\bs e_{\bar{\bs P}}}

\begin{document}
\title{
Decoherence in Neutrino Oscillation\\
between 3D Gaussian Wave Packets
}

\author{
Haruhi Mitani\thanks{E-mail: \tt mitani.phys@gmail.com} \mbox{} and
Kin-ya Oda\thanks{E-mail: \tt odakin@lab.twcu.ac.jp}\bigskip\\
\it\normalsize
Department of Mathematics, Tokyo Woman's Christian University, Tokyo 167-8585, Japan
}
\maketitle
\begin{abstract}\noindent
There is renewed attention to whether we can observe the decoherence effect in neutrino oscillation due to the separation of wave packets with different masses in near-future experiments. As a contribution to this endeavor, we extend the existing formulation based on a single 1D Gaussian wave function to an amplitude between two distinct 3D Gaussian wave packets, corresponding to the neutrinos being produced and detected, with different central momenta and spacetime positions and with different widths. We find that the spatial widths-squared for the production and detection appear additively in the (de)coherence length and in the localization factor for governing the propagation of the wave packet, whereas they appear as the reduced one (inverse of the sum of inverse) in the momentum conservation factor. The overall probability is governed by the ratio of the reduced to the sum.

\end{abstract}

\newpage
\section{Introduction}
It has been a quarter century since the existence of neutrino oscillation phenomena was established~\cite{Super-Kamiokande:1998kpq}.
However, many aspects, including the mass hierarchy, absolute mass values, and CP-violating phases, remain unresolved and await further investigation; see e.g.\ Ref.~\cite{ParticleDataGroup:2022pth} for a review.
At first, calculations of neutrino oscillation were conducted using plane waves~\cite{Pontecorvo:1957cp,Katayama:1962mx,Maki:1962mu}. Recently, more fundamental and precise discussions based on wave-packet formulations~\cite{Nussinov:1976uw, Bilenky:1978nj, Kayser:1981ye} have become prominent~\cite{Cardall:1999ze, Akhmedov:2008jn,Akhmedov:2009rb, Kopp:2009fa, Akhmedov:2010ua, Akhmedov:2010ms, Wu:2010yr, Akhmedov:2012uu, Ettefaghi:2020otb, Blasone:2021cau,Krueger:2023skk}.

The neutrino oscillation is suppressed by the decoherence effect due to wave-packet separation between different mass eigenstates beyond the so-called (de)coherence length~\cite{Giunti:1991ca,Giunti:1991sx,Giunti:1997wq,Giunti:2000kw,Beuthe:2002ej}; see also Ref.~\cite{Nussinov:1976uw} for an earlier ultra-relativistic expression for the coherence length.\footnote{
Another paper~\cite{Kayser:2010pr} is put on arXiv with caution: ``Below, we attach the incorrect first version of the article. It will be rewritten to explain the aforementioned problem and address related questions.''
}
(In Ref.~\cite{Ioannisian:1998ch}, the decoherence effect is pursued in a non-wavepacket approach, which is further developed recently~\cite{Karamitros:2022nnh}.)
In Refs.~\cite{deGouvea:2020hfl,deGouvea:2021uvg,Jones:2022hme}, decoherence due to wave packet separation is discussed with emphasis on experimental testability.
In the latest analysis~\cite{Jones:2022hme} among them, the density matrix for the produced neutrino state is traced out with respect to the accompanying charged lepton state to find a consistent decoherence effect to that in Ref.~\cite{Giunti:1997wq}, with a negative conclusion for the near-future detectability in reactor neutrino experiments such as JUNO.
A similar conclusion is made in a talk~\cite{Smirnov:2022aab}.

So far, all the above analyses are based on a single 1D Gaussian wave function, either in the position or momentum space, taking into account only the effect of the neutrino production process.
In this paper, we treat both the in and out neutrino states by 3D Gaussian wave packets to take into account the neutrino detection process too.

This Letter is organized as follows:
In Sec.~\ref{neutrino wave packet section}, we briefly review the Gaussian wave-packet formalism applied to the neutrino field.
In Sec.~\ref{decoherence section}, we show our main result: the amplitude and probability for the neutrino oscillation between two distinct Gaussian wave packets.
In Sec.~\ref{conclusion section}, we conclude this Letter.
In Appendix~\ref{saddle point section}, we present more details on the saddle-point approximation in deriving the amplitude between the Gaussian wave-packet states.
In Appendix~\ref{complete square with T}, we show a formula by first square-completing the exponent with respect to $T$.
In Appendix~\ref{two-flavor section}, we show a concrete two-flavor result.

\section{Neutrino wave packet}\label{neutrino wave packet section}
In this section, we first show how to treat the neutrino wave packet rigorously, and then present our approximation of neglecting the spin dependence. Readers who are more interested in phenomenological aspects may skip this section.

We work in $d=3$ spatial dimensions. The metric signature is taken to be $\pn{-,+,\dots,+}$. Greek indices $\mu,\nu,\dots$ run from 0 to $d$ and roman ones $i,j,\dots$ from 1 to $d$. We employ natural units $\hbar=c=1$ unless otherwise stated.
We do not use spacetime notation for momenta
and write $p:=\ab{\bs p}$, whereas the spacetime coordinates are written as $x:=\pn{x^0,\bs x}=\pn{x^0,x^1,\dots,x^d}$.

\subsection{Plane wave}
Let us first expand a free neutrino field in $I$th mass eigenstate (after the electroweak symmetry breaking) into the plane waves:
\al{
\wh\nu_I\fn{x}
	&=	\sum_s\int{\df^d\bs p\ov\pn{2\pi}^{d\ov2}}\sqbr{
		u_I\fn{\bs p,s}e^{-iE_I\fn{p}x^0+i\bs p\cdot\bs x}\,\wh a_I\fn{\bs p,s}
		+v_I\fn{\bs p,s}e^{iE_I\fn{p}x^0-i\bs p\cdot\bs x}\,\wh a^{\tx c\dagger}_I\fn{\bs p,s}
		},
}
where $E_I\fn{p}$ is the energy of the $I$th mass eigenstate,
$
E_I\fn{p}
	:=	\sqrt{m_I^2+p^2}
$,
and the annihilation and creation operators of the neutrino and antineutrino fields obey
\al{
\anticommutator{\wh a_I\fn{\bs p,s}}{\wh a_J^\dagger\fn{\bs p',s'}}
	&=	\delta_{IJ}\delta_{ss'}\delta^d\fn{\bs p-\bs p'}\wh1,\nn
\anticommutator{\wh a^\tx{c}_I\fn{\bs p,s}}{\wh a^{\tx c\dagger}_J\fn{\bs p',s'}}
	&=	\delta_{IJ}\delta_{ss'}\delta^d\fn{\bs p-\bs p'}\wh1,&
\tx{others}
	&=	0.\label{commutators of alpha}
}
If neutrinos are Majorana $\wh a^\tx{c}_I\fn{\bs p,s}=\wh a_I\fn{\bs p,s}$, the momentum-space plane-wave functions $u_I\fn{\bs p,s}$ and $v_I\fn{\bs p,s}$ are related by the Majorana condition accordingly.\footnote{
Here we adopt the notation in Ref.~\cite{Weinberg:1995mt}. The Majorana condition reads $v_I\fn{\bs p,s}=\pn{-\beta\ms C}u_I^*\fn{\bs p,s}$.
}
Hereafter, we assume the Majorana neutrino for simplicity; switching to a Dirac neutrino and/or inclusion of a sterile neutrino is straightforward.

A momentum eigenbasis of the free neutrino one-particle subspace can be defined by
\al{
\Ket{\bs p,s;I}
	&:=	\wh a_I^\dagger\fn{\bs p,s}\Ket{0}
}
such that the commutator~\eqref{commutators of alpha} leads to the normalization
\al{
\Braket{\bs p,s;I|\bs p,s';J}
	&=	\delta_{IJ}\delta_{ss'}\delta^d\fn{\bs p-\bs p'}.
	\label{plane-wave normalization}
}
In the one-particle subspace, the completeness relation (resolution of identity) is
\al{
\sum_{I,s}\int\df^d\bs p\Ket{\bs p,s;I}\Bra{\bs p,s;I}
	&=	\h1,
	\label{completeness}
}
where $\h1$ is the identity operator in the free neutrino one-particle subspace.

Finally, for a given momentum $\bs p$, we write the neutrino velocity in the $I$th mass eigenstate as
\al{
\bs v_I\fn{\bs p}
	:=	{\bs p\ov E_I\fn{p}}
	=	v_I\fn{p}\bs e_{\bs p},
	\label{neutrino velocity}
}
where $v_I\fn{p}:=\ab{\bs v_I\fn{\bs p}}$ and $\bs e_{\bs p}:=\bs p/p$.
Its direction does not depend on the mass eigenstates, whereas the magnitude does.

\subsection{Gaussian wave packet}
Here, we review the Gaussian wave-packet formalism, following Appendix~A in Ref.~\cite{Ishikawa:2018koj}, to spell out notations in this paper;
see also Refs.~\cite{Ishikawa:2020hph,Ishikawa:2021bzf} for its application in quantum field theory and Ref.~\cite{Oda:2021tiv} for a historical account of earlier works.

We define the Gaussian basis state $\Ket{X,\bs P,s;\sigma,I}$ in the free neutrino one-particle subspace by
\al{
\Braket{X,\bs P,s;\sigma,I|\bs p,s';J}
	&:=	\delta_{IJ}\delta_{ss'}\pn{\sigma\ov\pi}^{d/4}e^{-iE_I\fn{p}X^0+i\bs p\cdot\bs X-{\sigma\ov2}\pn{\bs p-\bs P}^2},
	\label{Gaussian basis defined}
}
which is centered at the spacetime point $X$ and the momentum $\bs P$, with $\sigma$ being its spatial width-squared.\footnote{
This definition actually breaks the Lorentz covariance~\cite{Oda:2021tiv}.
A manifestly Lorentz covariant formulation of a spinor wave packet can be found in Ref.~\cite{Oda:2023qek}.
}
Sandwiching by the momentum eigenbases, it is straightforward to show the completeness (resolution of identity) for the Gaussian basis states:
\al{
\sum_{I,s}\int{\df^d\bs X\,\df^d\bs P\ov\pn{2\pi}^d}\Ket{X,\bs P,s;\sigma,I}\Bra{X,\bs P,s;\sigma,I}
	&=	\h1.
	\label{completeness of Gaussian}
}
It is important that the completeness relation~\eqref{completeness of Gaussian} holds for arbitrary fixed $\sigma$ and $X^0$, which are not summed nor integrated.
Since the Gaussian wave packets form the complete set, they can be used to expand an arbitrary shape of a wave packet.

\subsection{Wave packet in interaction eigenstate}
In the standard three-flavor scheme (see e.g.\ Ref.~\cite{ParticleDataGroup:2022pth} for a review), a neutrino field in $\alpha$th interaction (flavor) eigenstate is rotated from the mass eigenstates by the so-to-say PKMTYMNS matrix~\cite{Pontecorvo:1957cp,Katayama:1962mx,Maki:1962mu}:
\al{
\wh\nu_\alpha\fn{x}
	&=	\sum_IU_{\alpha I}\wh\nu_I\fn{x},
}
where greek and upper-case roman letters $\alpha,\beta,\dots$ and $I,J,\dots$ run for $e,\mu,\tau$ and $1,2,3$, respectively.

It is important that what is mixed by the PKMTYMNS matrix $U_{\alpha I}$ is the neutrino field $\wh\nu_\alpha\fn{x}$ and is not the momentum eigenbasis $\Ket{\bs p,s;I}$.
% nor the position eigenbasis $\Ket{x,s;I}$.
Indeed the neutrino field in an interaction eigenstate, $\wh\nu_\alpha\fn{x}$, is related to the wave function in a mass eigenstate $\Ket{\bs p,s;I}$ by
\al{
\Bra{0}\wh\nu_\alpha\fn{x}\Ket{\bs p,s;I}
	&=	\sum_IU_{\alpha I}{u_I\fn{\bs p,s}e^{-iE_I\fn{p}x^0+i\bs p\cdot\bs x}\ov\pn{2\pi}^{d\ov2}},\\
\Bra{\bs p,s;I}\wh\nu_\alpha\fn{x}\Ket{0}
	&=	\sum_IU_{\alpha I}{v_I\fn{\bs p,s}e^{iE_I\fn{p}x^0-i\bs p\cdot\bs x}\ov\pn{2\pi}^{d\ov2}}.
}
In the literature, this point has been neglected so far. Including such a spinor effect is an interesting topic in itself and will be covered in a separate publication~\cite{OdaWada2}. For the purpose of the current study, we adhere to the convention of disregarding this effect and sloppily write a wave packet state in an interaction eigenstate $\nu_\alpha$ as
\al{
\Ket{X,\bs P;\sigma,\nu_\alpha}
	&=	\sum_IU_{\alpha I}^*\Ket{X,\bs P;\sigma,I}.
}

\section{Decoherence from wave-packet separation}\label{decoherence section}
\subsection{Oscillation probability of neutrino wave packet}
Now we compute the oscillation probability amplitude between interaction eigenstates of a neutrino $\nu_\alpha$ at a source and $\nu_\beta$ at a detector.
We take the initial and final states as Gaussian wave packets with spatial widths-squared $\sigma_\s$ and $\sigma_\D$.
These wave packets are centered at spacetime points $X_\s$ and $X_\D$ in the position space, as well as at $\bs P_\s$ and $\bs P_\D$ in the momentum space.
The amplitude reads
\al{
\Braket{\nu_\alpha ,\sigma_\s;X_\s,\bs P_\s|\nu_\beta ,\sigma_\D;X_\D,\bs P_\D}
%	&=	\sum_{I,J}U_{\beta J}U_{\alpha I}^*\Braket{J,\sigma_\D;X_\D,\bs P_\D|I,\sigma_\s;X_\s,\bs P_\s}\nn
	&=	\sum_IU_{\beta I}^*U_{\alpha I}\Braket{I,\sigma_\s;X_\s,\bs P_\s|I,\sigma_\D;X_\D,\bs P_\D},
	\label{amplitude}
}
where we have used the diagonality of the mass eigenstates~\eqref{plane-wave normalization} and \eqref{Gaussian basis defined}.

After expanding the amplitude~\eqref{amplitude} by the complete basis~\eqref{completeness}, we may compute it using the saddle-point approximation:
\al{
\Braket{\nu_\alpha ,\sigma_\s;X_\s,\bs P_\s|\nu_\beta ,\sigma_\D;X_\D,\bs P_\D}
	&\simeq
%		\pn{\sigma_\s\sigma_\D\ov\sigma_\tx{sum}^2}^{d\ov4}\sum_IU_{\beta I}U_{\alpha I}^*,\nn
%	&\quad\times
%			e^{
%			i\pn{\bar E_I-\bar{\bs P}\cdot\bar{\bs v}_I}T
%			-i\bar{\bs P}\cdot\pn{\bs L-\bar{\bs v}_IT}
%			-{1\ov2}\sigma_\tx{red}\pn{\bs P_\D-\bs P_\s}^2
%			-{\pn{\bs L-\bar{\bs v}_IT}^2\ov2\sigma_\tx{sum}}
%				}\nn
%	&=	
%		\sqbr{\sigma_\s\sigma_\D\ov\sigma_\tx{sum}^2}^{d\ov4}
		\pn{\sigma_\tx{red}\ov\sigma_\tx{sum}}^{d\ov4}
		e^{
			-i\bar{\bs P}\cdot\bs L
			-{\sigma_\tx{red}\ov2}\pn{\bs P_\D-\bs P_\s}^2
			}\nn
	&\quad\times
		\sum_IU_{\alpha I}U_{\beta I}^*
		{\exp\fnl{	
			i\bar E_IT
			-{\pn{\bs L-\bar{\bs v}_IT}^2\ov2\sigma_\tx{sum}}
				}
		\ov
			\pn{1-i{T\ov\sigma_\tx{sum}\bar E_I}}^{d-1\ov2}
			\sqrt{1-i{\pn{1-\bar v_I^2}T\ov\sigma_\tx{sum}\bar E_I}}
			},
				\label{starting amplitude}
}
where
\al{
\sigma_\tx{sum}
	&:=	\sigma_\s+\sigma_\D,&
\sigma_\tx{red}
	&:=	{\sigma_\s\sigma_\D\ov\sigma_\s+\sigma_\D},\\
T	&:=	X_\D^0-X_\s^0,&
\bs L
	&:=	\bs X_\D-\bs X_\s,&
\bar{\bs P}
	&:=	{\sigma_\s\bs P_\s+\sigma_\D\bs P_\D\ov\sigma_\s+\sigma_\D},\\
\bar E_I
	&:=	E_I\fn{\bar P},&
\bar{\bs v}_I
	&:=	\bs v_I\fn{\bar{\bs P}},
}
in which $\bar P:=\ab{\bar{\bs P}}$; see Appendix~\ref{saddle point section} for more details on the saddle-point approximation in deriving this amplitude.
The following relation is also useful:
\al{
\bar{\bs v}_I
	=	\bar v_I\ePbar ,\label{parallelness of velocities}
}
where
$\bar v_I:=\ab{\bar{\bs v}_I}=\bar P/E_I\fn{\bar P}$; see Eq.~\eqref{neutrino velocity}.
%Physically,
%$\sigma_\tx{sum}$ and $\sigma_\tx{red}$ are the summed and reduced width-squared, respectively;
%$T$ is the elapsed time;
%$\bs L$ is the displacement from the source to the detector;
%$\bar{\bs P}$ is the weighted average of the initial and final momenta; and
%$\bar E_I$ and $\bar{\bs v}_I$ are, respectively, the on-shell energy and velocity of the $I$th mass eigenstate for the weight-averaged momentum.
In physical terms, $\sigma_\tx{sum}$ and $\sigma_\tx{red}$ represent the summed and reduced width-squared, respectively. Additionally, $T$ denotes the elapsed time, while $\bs L$ indicates the displacement from the source to the detector. The weighted average of the initial and final momenta is symbolized by $\bar{\bs P}$. Finally, $\bar E_I$ and $\bar{\bs v}_I$ correspond to the on-shell energy and velocity of the $I$th mass eigenstate for the weight-averaged momentum, respectively. 

In the denominator in Eq.~\eqref{starting amplitude}, the factors $\pn{\cdots}^{d-1\ov2}$ and $\sqrt{\cdots}$ correspond to the broadening with time of the wave packet in the transverse and longitudinal spatial directions, respectively, as observed in the Gaussian wave function within the position space~\cite{Ishikawa:2005zc,Ishikawa:2018koj}, where only the transverse directions broaden with time in the ultra-relativistic limit.\footnote{
For large $T$ ($\sim L$) ($\gg\sigma_\tx{sum}\bar E_I$), the former factor $\pn{\cdots}^{d-1\ov2}$ (in $d=3$) gives the suppression $L^{-1}$ for the amplitude, representing the geometric suppression factor due to the Gauss law, favorably comparing with the QFT findings in Refs.~\cite{{Ioannisian:1998ch,Karamitros:2022nnh}}.
}
Hereafter, we will write them
$C_I:=\pn{1-i{T\ov\sigma_\tx{sum}\bar E_I}}^{d-1\ov2}\sqrt{1-i{\pn{1-\bar v_I^2}T\ov\sigma_\tx{sum}\bar E_I}}$.

The oscillation probability now reads
\al{
P\fn{\alpha\to\beta}
	&=	\Ab{\Braket{\nu_\alpha ,\sigma_\s;X_\s,\bs P_\s|\nu_\beta ,\sigma_\D;X_\D,\bs P_\D}}^2\nn
	&=	
		\sum_{I,J}U_{\alpha I}U_{\beta I}^*U_{\alpha J}^*U_{\beta J}e^{i\pn{\bar E_I-\bar E_J}T}\nn
	&\quad\times
		\pn{\sigma_\tx{red}\ov\sigma_\tx{sum}}^{d\ov2}
		{\exp\fnl{
			-\sigma_\tx{red}\pn{\bs P_\D-\bs P_\s}^2
			-{\pn{\bs L-\bar{\bs v}_IT}^2\ov2\sigma_\tx{sum}}
			-{\pn{\bs L-\bar{\bs v}_JT}^2\ov2\sigma_\tx{sum}}
				}
		\ov
		C_IC_J^*
		}.
%	&=	
%		\sum_{I,J}U_{\alpha I}U_{\beta I}%*U_{\alpha J}%*U_{\beta J}e%{i\pn{\bar E_I-\bar E_J}T}\nn
%	&\quad\times
%		\pn{\sigma_\tx{red}\ov\sigma_\tx{sum}}%{d\ov2}
%		{\exp\fnl{
%			-\sigma_\tx{red}\pn{\bs P_\D-\bs P_\s}%2
%			-{\pn{\bs L-\bar{\bs v}_IT}%2\ov2\sigma_\tx{sum}}
%			-{\pn{\bs L-\bar{\bs v}_JT}%2\ov2\sigma_\tx{sum}}
%				}
%		\ov
%		\pn{1
%			+{T%2\ov\sigma_\tx{sum}%2\bar E_I\bar E_J}
%			+i{\pn{\bar E_I-\bar E_J}T\ov\sigma_\tx{sum}\bar E_I\bar E_J}
%			}%{d-1\ov2}
%		\sqrt{
%			1
%			+{\pn{1-\bar v_I%2}\pn{1-\bar v_J%2}T%2\ov\sigma_\tx{sum}%2\bar E_I\bar E_J}
%			-i{T\ov\sigma_\tx{sum}}\pn{{1-\bar v_I%2\ov\bar E_I}-{1-\bar v_J%2\ov\bar E_J}}
%			}
%		}.
				\label{master formula}
}
We emphasize that what we have computed in Eq.~\eqref{starting amplitude} is the quantum mechanical amplitude between normalizable states whose absolute square directly becomes the probability~\eqref{master formula}.
The second-last line in Eq.~\eqref{master formula} is nothing but the ordinary plane-wave oscillation probability.
Physically, each term in the $I,J$ summation can be regarded as the interference between the $I$th and $J$th mass eigenstates, both acting as mediators between the initial interaction eigenstate $\nu_\alpha$ and the final $\nu_\beta$.

The last line exhibits the wave-packet effects: The first term in the exponent shows how the deviation from the momentum conservation is exponentially suppressed. In a plane-wave limit $\sigma_\tx{red}\to\infty$, this part is reduced to the momentum delta function:
\al{
\sigma_\tx{red}^{d\ov2}e^{-\sigma_\tx{red}\pn{\bs P_\D-\bs P_\s}^2}
	&\to	\pi^{d\ov2}\delta^d\fn{\bs P_\D-\bs P_\s}.
}
Meanwhile, the second and last terms exhibit the localization of the wave packets at $\bs L=\bar{\bs v}_IT$ and $\bar{\bs v}_JT$ for the $I$th and $J$th mass eigenstates, respectively.

\subsection{Decoherence effect}
Here we show how the decoherence effect appears in two different approaches.
Firstly, we square-complete the real part of the exponent, which represents the wave-packet effects, with respect to $\bs L$:
\al{
P\fn{\alpha\to\beta}
	&=	
		\pn{\sigma_\tx{red}\ov\sigma_\tx{sum}}^{d\ov2}
		e^{-\sigma_\tx{red}\pn{\bs P_\D-\bs P_\s}^2}
		\sum_{I,J}U_{\alpha I}U_{\beta I}^*U_{\alpha J}^*U_{\beta J}
			e^{i\pn{\bar E_I-\bar E_J}T}\nn
	&\quad\times
		{\exp\fnl{
			-{\bs L_\perp^2\ov\sigma_\tx{sum}}
			-{\pn{L_\parallel-{\bar v_I+\bar v_J\ov2}T}^2\ov\sigma_\tx{sum}}
			-{\pn{\bar v_I-\bar v_J}^2\ov4\sigma_\tx{sum}}T^2
				}
		\ov
		C_IC_J^*
		},
				\label{L square completed}
}
where we have used the parallelism of the velocities~\eqref{parallelness of velocities} and have defined\footnote{
Configurations of $\bs L\cdot\ePbar<0$ are exponentially suppressed for $T>0$, and are disregarded here.
}
\al{
\bs L_\parallel
	&:=	\pn{\bs L\cdot\ePbar}\ePbar,&
L_\parallel
	&:=	\ab{\bs L_\parallel},&
\bs L_\perp
	&:=	\bs L-\bs L_\parallel.
}
The last term in the exponent (in the last line) can be regarded as the decoherence effect due to the separation of the wave packets.
That is, the interference between $I$th and $J$th mass eigenstates is exponentially suppressed when the elapsed time exceeds a certain value, $T\gtrsim T_{\tx{coh}\,IJ}$, where
\al{
T_{\tx{coh}\,IJ}
	&:=	2{\sqrt{\sigma_\tx{sum}}\ov\ab{\bar v_I-\bar v_J}}.
}
Note that the elapsed time $T$ corresponds to the average length of the propagations of the $I$th and $J$th neutrino wave packets because of the second-last term in the exponent, which localizes the probability around
\al{
L_\parallel={\bar v_I+\bar v_J\ov2}T
	\approx	T,
}
where $\approx$ denotes the ultra-relativistic limit, which gives
\al{
T_{\tx{coh}\,IJ}
	&\approx
		{4\bar P^2\sqrt{\sigma_\tx{sum}}\ov\ab{m_I^2-m_J^2}}.
}

Let us turn to the second approach.
The exact timings of the neutrino emission and detection at the source and detector, respectively, are hardly measured in actual experiments, particularly in the case of reactor neutrinos.
Therefore, for developing physical insight, it would be helpful to square-complete $T$ first. The result is
\al{
P\fn{\alpha\to\beta}
	&=	\pn{\sigma_\tx{red}\ov\sigma_\tx{sum}}^{d\ov2}
		e^{-\sigma_\tx{red}\pn{\bs P_\D-\bs P_\s}^2}
		\sum_{I,J}U_{\alpha I}U_{\beta I}^*U_{\alpha J}^*U_{\beta J}
			e^{i\pn{\bar E_I-\bar E_J}T}\nn
	&\quad\times
		{\exp\fnl{
			-{\bs L_\perp^2\ov\sigma_\tx{sum}}
			-{\pn{\bar v_I^2+\bar v_J^2}
				\pn{
					T-{\bar v_I+\bar v_J\ov\bar v_I^2+\bar v_J^2}L_\parallel
					}^2
				\ov2\sigma_\tx{sum}}
		-{{\pn{\bar v_I-\bar v_J}^2\ov\bar v_I^2+\bar v_J^2}L_\parallel^2\ov2\sigma_\tx{sum}}
				}
		\ov
		C_IC_J^*
		},
				\label{T partly square-completed}
}

Several comments are in order:
\begin{itemize}
\item From the last term in the exponent in Eq.~\eqref{T partly square-completed}, we again observe the emergence of a coherence length $L_{\tx{coh}\,IJ}$:
\al{
L_{\tx{coh}\,IJ}
	&:=	\sqrt{2\pn{\bar v_I^2+\bar v_J^2}}
			{\sqrt{\sigma_\tx{sum}}\ov\ab{\bar v_I-\bar v_J}}.
}
That is, the interference between $I$th and $J$th mass eigenstates is exponentially suppressed when $L_\parallel\gtrsim L_{\tx{coh}\,IJ}$.
We see that the first and second approaches agree:
\al{
L_{\tx{coh}\,IJ}
	\approx T_{\tx{coh}\,IJ}
	\approx {4\bar P^2\sqrt{\sigma_\tx{sum}}\ov m_I^2-m_J^2}.
	\label{decoherence length and time}
}
%since neutrinos are ultra-relativistic, $\bar v_I^2\approx\bar v_J^2\approx1$.
%This expression agrees with the classic results~\cite{Nussinov:1976uw,Giunti:1991ca}. In particular, the (de)coherence length becomes infinity in the degenerate-mass limit $m_I=m_J$, which is consistent with the disappearance of the exponential suppression factor in Eqs.~\eqref{L square completed} and \eqref{T partly square-completed} due to $\bar v_I=\bar v_J$ in the degenerate-mass limit.
\item The second-last term effectively chooses configurations around
\al{
T	&=	{\bar v_I+\bar v_J\ov\bar v_I^2+\bar v_J^2}L_\parallel
	\approx L_\parallel.
}
This backs up the equivalence of two approaches~\eqref{decoherence length and time}.
\item The third-last term shows that the interference between $I$th and $J$th mass eigenstates is exponentially suppressed when the size of perpendicular displacement $\ab{\bs L_\perp}$ exceeds the square-summed wave packet size $\sqrt{\sigma_\tx{sum}}$.
\end{itemize}

\subsection{Ultra-relativistic limit}
For completeness, we also write down the ultra-relativistic limit of Eq.~\eqref{L square completed} and \eqref{T partly square-completed}, where the dots denote terms of $\mc O\fn{\bar P^{-3}}$:
\al{
P\fn{\alpha\to\beta}
	&=
		e^{-\sigma_\tx{red}\pn{\bs P_\D-\bs P_\s}^2}
		\sum_{I,J}U_{\alpha I}U_{\beta I}^*U_{\alpha J}^*U_{\beta J}
			e^{i{m_I^2-m_J^2\ov2\bar P}T\pn{1-{m_I^2+m_J^2\ov4\bar P^2}+\cdots}}\nn
	&\quad\times
		\pn{\sigma_\tx{red}\ov\sigma_\tx{sum}}^{d\ov2}
		{\exp\Bigg[
			-{\bs L_\perp^2\ov\sigma_\tx{sum}}
			-{\pn{L_\parallel-T}^2\ov\sigma_\tx{sum}}
			-{\pn{m_I^2+m_J^2}\pn{L_\parallel-T}T\ov2\sigma_\tx{sum}\bar P^2}
%			+{\pn{m_I^4+m_J^4}\pn{3L_\parallel-4T}T\ov8\sigma_\tx{sum}\bar P^4}
		+\cdots
			\Bigg]
		\ov
		\pn{1+{T^2\ov\sigma_\tx{sum}^2\bar P^2}+\cdots}^{d-1\ov2}
		\sqrt{1+\cdots}
		}.
			\label{another expansion}
}
\begin{comment}
We point out that if we manage to have an experiment to adjust the timing $T$ at $L_\parallel$, the second-last line disappears, and the decoherence time from the last line becomes\footnote{
This point depends on how precisely one can set up a real experiment.
In Eqs.~\eqref{L square completed} and \eqref{T partly square-completed}, the wave-packet center is supposed to be localized at the timing $T=L_\parallel\bigl(1+{m_I^2+m_J^2\ov4\bar{\bs P}^2}-{m_I^4-m_I^2m_J^2+m_J^4\ov8\bar{\bs P}^4}+\cdots\bigr)$ and $L_\parallel\bigl(1+{m_I^2+m_J^2\ov4\bar{\bs P}^2}-{3m_I^4-4m_I^2m_J^2+3m_J^4\ov16\bar{\bs P}^4}+\cdots\bigr)$, respectively.
That is, the supposed center of localization is not at $T=L_\parallel$ in both. In this sense, one might regard the length~\eqref{artificial decoherence length} as an artifact of expanding the probability around (or performing an experiment at) the ``wrong'' timing $T=L_\parallel$ as in Eq.~\eqref{another expansion}.
}
\al{
{2\sqrt2\bar P^2\sqrt{\sigma_\tx{sum}}\ov\sqrt{m_I^4+m_J^4}},
	\label{artificial decoherence length}
}
which can be significantly shorter than Eq.~\eqref{decoherence length and time} for a degenerate (or inverted hierarchy) scenario, $m_I^2-m_J^2\ll m_I^2+m_J^2$, though the scenario itself is already seriously constrained.
\end{comment}

\section{Conclusion}\label{conclusion section}
In this Letter, we computed the amplitude and probability for the neutrino oscillation between two distinct Gaussian wave packets that have different centers of position and momentum as well as different widths.

We find that the spatial widths-squared for the production and detection appear additively in the (de)coherence length as well as in the localization factor governing the propagation of the wave packet, whereas they appear as the reduced one (inverse of the sum of inverse) in the momentum conservation factor. The overall probability is governed by the ratio of the reduced to the sum.
We have obtained the (de)coherence lengths in two ways, which coincide with each other in the ultra-relativistic limit.

\subsection*{Acknowledgement}
We thank Juntaro Wada for the useful discussion and comments.
The work of K.O.\ is in part supported by JSPS KAKENHI Grant Nos.~19H01899 and 21H01107.

\appendix
\section*{Appendix}

\section{Saddle-point approximation}\label{saddle point section}
The amplitude~\eqref{starting amplitude} is the leading result of the saddle-point approximation complemented by the large spatial-width expansion, along the lines of Appendix~B in Ref.~\cite{Ishikawa:2018koj}. For completeness, we present both the pre-expansion saddle-point result and the next leading terms.

The saddle-point approximation yields
\al{
\Braket{\nu_\alpha ,\sigma_\s;X_\s,\bs P_\s|\nu_\beta ,\sigma_\D;X_\D,\bs P_\D}
	&=	\pn{\sigma_\s\ov\pi}^{d\ov4}\pn{\sigma_\D\ov\pi}^{d\ov4}
		\sum_IU_{\alpha I}U_{\beta I}^*\int\df^d\bs p\,e^{G_I\pn{\bs p}}\nn
	&\simeq
		\pn{\sigma_\tx{red}\ov\sigma_\tx{sum}}^{d\ov4}
		\sum_IU_{\alpha I}U_{\beta I}^*\nn
	&\quad\times
		{e^{
			G_I\pn{\bs p_\star}
			}
		\ov
		\pn{1-i{T\ov\sigma_\tx{sum}E_I\fn{\bs p_\star}}}^{d-1\ov2}
		\sqrt{1-i\pn{1-\bs v_I^2\fn{\bs p_\star}}{T\ov\sigma_\tx{sum}E_I\fn{\bs p_\star}}}
		},
			\label{saddle-point result}
}
where
\al{
G_I\fn{\bs p}
	&:=	iE_I\fn{\bs p}T-i\bs p\cdot\bs L
		\ub{
			-{\sigma_s\ov2}\pn{\bs p-\bs P_\s}^2-{\sigma_\D\ov2}\pn{\bs p-\bs P_\D}^2
			}_{
				-{\sigma_\tx{sum}\ov2}\pn{\bs p-\bar{\bs P}}^2
				-{\sigma_\tx{red}\ov2}\pn{\bs P_\D-\bs P_\s}^2
				}
}
is the exponent before performing the $\bs p$-integration;
$\bs p_\star$ is the solution to the saddle-point equation $\p G_I\fn{\bs p_\star}/\p p^i=0$;
and, around the saddle point $\bs p_\star$, the exponent is approximated as usual: $G_I\fn{\bs p}\simeq G_I\fn{\bs p_\star}+{1\ov2}{\p^2G_I\fn{\bs p_\star}\ov\p p^i\p p^j}\pn{p^i-p_\star^i}\Pn{p^j-p_\star^j}$. Concretely,
\al{
{\p G_I\fn{\bs p}\ov\p p^i}
	&=	-i\pn{L-v_I^i\fn{\bs p}T}-\sigma_\tx{sum}\pn{p^i-\bar P^i},\\
{\p^2G_I\fn{\bs p}\ov\p p^i\p p^j}
	&=	-\sigma_\tx{sum}\pn{\delta^{ij}-i{\delta^{ij}-v_I^i\fn{\bs p}v_I^j\fn{\bs p}\ov\sigma_\tx{sum}E_I\fn{\bs p}}T}.
	\label{second derivative}
}

Iteratively solving $\bs p_\star=\sum_{n=0}^\infty\bs p_{(n)}$ with $\bs p_{(n)}=\mc O\fn{\sigma_\tx{sum}^{-n}}$, we may obtain arbitrary higher order terms. For example, the next-leading order term of the saddle point is
\al{
\bs p_\star
	&=	\bar{\bs P}
		-i{\bs L-\bar{\bs v}_IT\ov\sigma_\tx{sum}}
		+{\pn{\bs L-\bar{\bs v}_IT}-\bar{\bs v}_I\Pn{\bar{\bs v}_I\cdot\pn{\bs L-\bar{\bs v}_IT}}\ov\sigma_\tx{sum}^2\bar E_I}T
		+\mc O\fn{\sigma_\tx{sum}^{-3}}.
		\label{pstar given}
}
%We see that the Gaussian integral~\eqref{saddle-point result} is always possible because the real part of the second derivative~\eqref{second derivative} is always negative at the saddle point~\eqref{pstar given}, whenever the large $\sigma_\tx{sum}$ expansion is valid.

\section{Complete square with respect to $T$}\label{complete square with T}
As discussed in the main text, $T$ is hardly measured. It would be useful to obtain an expression of square-completed $T$ from all the exponents in Eq.~\eqref{master formula} in order to integrate out $T$.\footnote{\label{emergence of integral}
Our analysis in this paper is within the level of quantum mechanics. In quantum field theory, an example of the emergence of such integration over $T$ from the final-state phase-space integral can be seen in e.g.\ Sec.~4 in Ref.~\cite{Ishikawa:2018koj}.
}
The result is
\al{
P\fn{\alpha\to\beta}
	&=	
		\pn{\sigma_\tx{red}\ov\sigma_\tx{sum}}^{d\ov2}
		e^{-\sigma_\tx{red}\pn{\bs P_\D-\bs P_\s}^2}
		\sum_{I,J}
			U_{\alpha I}U_{\beta I}^*U_{\alpha J}^*U_{\beta J}
			\nn
	&\quad\times
		{\exp\sqbr{
		-{\bar v_I^2+\bar v_J^2\ov2\sigma_\tx{sum}}
			\pn{
			T-T_{\fs\,IJ}
			}^2
		-\sigma_\tx{sum}\pn{\bar E_I-\bar E_J}^2
		-{\bs L_\perp^2\ov\sigma_\tx{sum}}
		-{\pn{\bar v_I-\bar v_J}^2
			\sqbr{
			L_\parallel
			+i{\sigma_\tx{sum}\pn{\bar v_I+\bar v_J}\pn{\bar E_I-\bar E_J}\ov
				\pn{\bar v_I-\bar v_J}^2}
			}^2
			\ov2\pn{\bar v_I^2+\bar v_J^2}\sigma_\tx{sum}}
			}\ov
			C_IC_J^*
			},
}
where we have defined a complex parameter:
\al{
T_{\fs\,IJ}
	&:=	{L_\parallel\pn{\bar v_I+\bar v_J}
				+i\sigma_\tx{sum}\pn{\bar E_I-\bar E_J}\ov\bar v_I^2+\bar v_J^2}.
}
Its further analysis will be presented elsewhere; see footnote~\ref{emergence of integral}.

In a heuristic ultra-relativistic limit leaving only characteristic leading contributions in each part, this reduces to
\al{
P\fn{\alpha\to\beta}
	&\approx
		\pn{\sigma_\tx{red}\ov\sigma_\tx{sum}}^{d\ov2}
		{e^{-\sigma_\tx{red}\pn{\bs P_\D-\bs P_\s}^2}
		\ov
		\pn{1+{T^2\ov\sigma_\tx{sum}^2\bar P^2}}^{d-1\ov2}
		}
		\sum_{I,J}U_{\alpha I}U_{\beta I}^*U_{\alpha J}^*U_{\beta J}\nn
	&\quad\times\exp\!\Bigg\{
		-{\pn{T-T_{\fs\,IJ}}^2\ov\sigma_\tx{sum}}
		-\sigma_\tx{sum}{\pn{m_I^2-m_J^2}^2\ov4\bar P^4}\nn
	&\phantom{\quad\times\exp\!\Bigg\{}
		-{\bs L_\perp^2\ov\sigma_\tx{sum}}
		-{\pn{m_I^2-m_J^2}^2
			\sqbr{
			L_\parallel
			+4i{\bar P^3\sigma_\tx{sum}\ov
				{m_I^2-m_J^2}}
			}^2
			\ov16\bar P^4\sigma_\tx{sum}}
			\Bigg\},
}
with
\al{
T_{\fs\,IJ}
	&\approx
		L_\parallel
				+i\sigma_\tx{sum}{m_I^2-m_J^2\ov4\bar P}.
}

\section{Two-flavor example}\label{two-flavor section}
It might be instructive to present a concrete two-flavor example between interaction eigenstates, say, $\nu_\alpha$ and $\nu_\beta$, with the PKMTYMNS matrix
\al{
%\bmat{U_{\alpha I}}_{\alpha=a,b;I=1,2}
\bmat{U_{\alpha1}&U_{\alpha2}\\ U_{\beta1}&U_{\beta2}}
	&=	\bmat{\cos\theta&\sin\theta\\-\sin\theta&\cos\theta}.
}
The resultant probability of non-oscillation from Eq.~\eqref{L square completed} is
\al{
P\fn{\alpha\to\alpha}
	&=	\pn{\sigma_\tx{red}\ov\sigma_\tx{sum}}^{d\ov2}
		e^{-\sigma_\tx{red}\pn{\bs P_\D-\bs P_\s}^2}
			\sum_{I,J}\ab{U_{\alpha I}}^2\ab{U_{\alpha J}}^2
		{e^{i\pn{\bar E_I-\bar E_J}T
				-{\bs L_\perp^2\ov\sigma_\tx{sum}}
				-{\pn{L_\parallel-{\bar v_I+\bar v_J\ov2}T}^2\ov\sigma_\tx{sum}}
				-{\pn{\bar v_I-\bar v_J}^2\ov4\sigma_\tx{sum}}T^2
					}
			\ov
				C_IC_J^*
			}\nn
	&=	\pn{\sigma_\tx{red}\ov\sigma_\tx{sum}}^{d\ov2}
		e^{
			-\sigma_\tx{red}\pn{\bs P_\D-\bs P_\s}^2
			-{\bs L_\perp^2\ov\sigma_\tx{sum}}
			}\nn
	&\quad\times\Bigg(
			\cos^4\theta\,{e^{-{\pn{L_\parallel-\bar v_1T}^2\ov\sigma_\tx{sum}}}\ov\ab{C_1}^2}
			+\sin^4\theta\,{e^{-{\pn{L_\parallel-\bar v_2T}^2\ov\sigma_\tx{sum}}}\ov\ab{C_2}^2}
			\nn
	&\phantom{\quad\times\Bigg(}
				+2\cos\fnl{\pn{\bar E_1-\bar E_2}T-\Delta_{12}}\cos^2\theta\sin^2\theta\,
				{e^{
					-{\pn{L_\parallel-{\bar v_1+\bar v_2\ov2}T}^2\ov\sigma_\tx{sum}}
					-{\pn{\bar v_1-\bar v_2}^2\ov4\sigma_\tx{sum}}T^2
					}
				\ov\ab{C_1C_2}}
			\Bigg),
}
where
$\Delta_{12}
	:=	{d-1\ov2}\arg\fn{1-{iT\ov\sigma_\tx{sum}\bar E_1}}
		+{d-1\ov2}\arg\fn{1+{iT\ov\sigma_\tx{sum}\bar E_2}}
		+{1\ov2}\arg\fn{1-{i\pn{1-\bar v_1^2}T\ov\sigma_\tx{sum}\bar E_1}}
		+{1\ov2}\arg\fn{1+{i\pn{1-\bar v_2^2}T\ov\sigma_\tx{sum}\bar E_2}}$,
and the one from Eq.~\eqref{T partly square-completed} is
\al{
P\fn{\alpha\to\alpha}
%	&=	\pn{\sigma_\tx{red}\ov\sigma_\tx{sum}}^{d\ov2}
%		e^{-\sigma_\tx{red}\pn{\bs P_\D-\bs P_\s}^2}
%		\sum_{I,J}\ab{U_{aI}}^2\ab{U_{aJ}}^2e^{-i\pn{\bar E_I-\bar E_J}T}\nn
%	&\quad\times
%		\exp\fnl{
%			-{\pn{\bar v_I^2+\bar v_J^2}
%				\pn{
%					T-{\bs L_\parallel\cdot\pn{\bar{\bs v}_I+\bar{\bs v}_J}\ov\bar v_I^2+\bar v_J^2}
%					}^2
%				\ov2\sigma_\tx{sum}}
%		-{\bs L_\perp^2\ov\sigma_\tx{sum}}
%		-{\pn{\bar{\bs v}_I-\bar{\bs v}_J}^2\ov2\pn{\bar v_I^2+\bar v_J^2}\sigma_\tx{sum}}L_\parallel^2
%				}\nn
	&=	\pn{\sigma_\tx{red}\ov\sigma_\tx{sum}}^{d\ov2}
		e^{-\sigma_\tx{red}\pn{\bs P_\D-\bs P_\s}^2-{\bs L_\perp^2\ov\sigma_\tx{sum}}}\nn
	&\quad\times\Bigg(
			\cos^4\theta\,{e^{
				-{\pn{L_\parallel-\bar v_1T}^2\ov\sigma_\tx{sum}}
				}\ov\ab{C_1}^2}
			+\sin^4\theta\,{e^{
				-{\pn{L_\parallel-\bar v_2T}^2\ov\sigma_\tx{sum}}
				}\ov\ab{C_2}^2}\nn
	&\phantom{\quad\times\Bigg(}
			+2\cos\fnl{\pn{\bar E_1-\bar E_2}T-\Delta_{12}}
				\cos^2\theta\sin^2\theta\,
			{e^{
					-{\pn{\bar v_1+\bar v_2}^2
						\pn{
						L_\parallel-{\bar v_1^2+\bar v_2^2\ov\bar v_1+\bar v_2}T
						}^2
						\ov2\pn{\bar v_1^2+\bar v_2^2}\sigma_\tx{sum}}
					-{\pn{\bar v_1-\bar v_2}^2
						\ov2\pn{\bar v_1^2+\bar v_2^2}\sigma_\tx{sum}}L_\parallel^2
					}\ov \ab{C_1C_2}}
			\Bigg).
}

In the ultra-relativistic limit, they both reduce to
\al{
P\fn{\alpha\to\alpha}
	&\approx
		\pn{\sigma_\tx{red}\ov\sigma_\tx{sum}}^{d\ov2}
		{e^{
			-\sigma_\tx{red}\pn{\bs P_\D-\bs P_\s}^2
			-{\bs L_\perp^2\ov\sigma_\tx{sum}}
			-{\pn{L_\parallel-T}^2\ov\sigma_\tx{sum}}}
			\ov
			\pn{1+{T^2\ov\sigma_\tx{sum}^2\bar P^2}}^{d-1\ov2}
			}\nn
	&\quad\times
	\Bigg(
			\cos^4\theta\,e^{
				-{m_1^2\pn{L_\parallel-T}T\ov\sigma_\tx{sum}\bar P^2}
				}
			+\sin^4\theta\,e^{
				-{m_2^2\pn{L_\parallel-T}T\ov\sigma_\tx{sum}\bar P^2}
				}\nn
	&\phantom{\quad\times\Bigg(}
			+2\cos\fnl{\pn{\bar E_1-\bar E_2}T}
					\cos^2\theta\sin^2\theta\,e^{
				-{\pn{m_1^2+m_2^2}\pn{L_\parallel-T}T\ov2\sigma_\tx{sum}\bar P^2}
				}
		\Bigg).
}

\bibliographystyle{JHEP}
\bibliography{refs}

\providecommand{\href}[2]{#2}\begingroup\raggedright\begin{thebibliography}{10}

\bibitem{Super-Kamiokande:1998kpq}
{\scshape Super-Kamiokande} collaboration, \emph{{Evidence for oscillation of
  atmospheric neutrinos}},
  \href{https://doi.org/10.1103/PhysRevLett.81.1562}{\emph{Phys. Rev. Lett.}
  {\bfseries 81} (1998) 1562}
  [\href{https://arxiv.org/abs/hep-ex/9807003}{{\ttfamily hep-ex/9807003}}].

\bibitem{ParticleDataGroup:2022pth}
{\scshape Particle Data Group} collaboration, \emph{{Review of Particle
  Physics}}, \href{https://doi.org/10.1093/ptep/ptac097}{\emph{PTEP} {\bfseries
  2022} (2022) 083C01}.

\bibitem{Pontecorvo:1957cp}
B.~Pontecorvo, \emph{{Mesonium and anti-mesonium}}, {\emph{Sov. Phys. JETP}
  {\bfseries 6} (1957) 429}.

\bibitem{Katayama:1962mx}
Y.~Katayama, K.~Matumoto, S.~Tanaka and E.~Yamada, \emph{{Possible unified
  models of elementary particles with two neutrinos}},
  \href{https://doi.org/10.1143/PTP.28.675}{\emph{Prog. Theor. Phys.}
  {\bfseries 28} (1962) 675}.

\bibitem{Maki:1962mu}
Z.~Maki, M.~Nakagawa and S.~Sakata, \emph{{Remarks on the unified model of
  elementary particles}}, \href{https://doi.org/10.1143/PTP.28.870}{\emph{Prog.
  Theor. Phys.} {\bfseries 28} (1962) 870}.

\bibitem{Nussinov:1976uw}
S.~Nussinov, \emph{{Solar Neutrinos and Neutrino Mixing}},
  \href{https://doi.org/10.1016/0370-2693(76)90648-1}{\emph{Phys. Lett. B}
  {\bfseries 63} (1976) 201}.

\bibitem{Bilenky:1978nj}
S.M.~Bilenky and B.~Pontecorvo, \emph{{Lepton Mixing and Neutrino
  Oscillations}},
  \href{https://doi.org/10.1016/0370-1573(78)90095-9}{\emph{Phys. Rept.}
  {\bfseries 41} (1978) 225}.

\bibitem{Kayser:1981ye}
B.~Kayser, \emph{{On the Quantum Mechanics of Neutrino Oscillation}},
  \href{https://doi.org/10.1103/PhysRevD.24.110}{\emph{Phys. Rev. D} {\bfseries
  24} (1981) 110}.

\bibitem{Cardall:1999ze}
C.Y.~Cardall, \emph{{Coherence of neutrino flavor mixing in quantum field
  theory}}, \href{https://doi.org/10.1103/PhysRevD.61.073006}{\emph{Phys. Rev.
  D} {\bfseries 61} (2000) 073006}
  [\href{https://arxiv.org/abs/hep-ph/9909332}{{\ttfamily hep-ph/9909332}}].

\bibitem{Akhmedov:2008jn}
E.K.~Akhmedov, J.~Kopp and M.~Lindner, \emph{{Oscillations of Mossbauer
  neutrinos}}, \href{https://doi.org/10.1088/1126-6708/2008/05/005}{\emph{JHEP}
  {\bfseries 05} (2008) 005} [\href{https://arxiv.org/abs/0802.2513}{{\ttfamily
  0802.2513}}].

\bibitem{Akhmedov:2009rb}
E.K.~Akhmedov and A.Y.~Smirnov, \emph{{Paradoxes of neutrino oscillations}},
  \href{https://doi.org/10.1134/S1063778809080122}{\emph{Phys. Atom. Nucl.}
  {\bfseries 72} (2009) 1363}
  [\href{https://arxiv.org/abs/0905.1903}{{\ttfamily 0905.1903}}].

\bibitem{Kopp:2009fa}
J.~Kopp, \emph{{Mossbauer neutrinos in quantum mechanics and quantum field
  theory}}, \href{https://doi.org/10.1088/1126-6708/2009/06/049}{\emph{JHEP}
  {\bfseries 06} (2009) 049} [\href{https://arxiv.org/abs/0904.4346}{{\ttfamily
  0904.4346}}].

\bibitem{Akhmedov:2010ua}
E.K.~Akhmedov and A.Y.~Smirnov, \emph{{Neutrino oscillations: Entanglement,
  energy-momentum conservation and QFT}},
  \href{https://doi.org/10.1007/s10701-011-9545-4}{\emph{Found. Phys.}
  {\bfseries 41} (2011) 1279}
  [\href{https://arxiv.org/abs/1008.2077}{{\ttfamily 1008.2077}}].

\bibitem{Akhmedov:2010ms}
E.K.~Akhmedov and J.~Kopp, \emph{{Neutrino Oscillations: Quantum Mechanics vs.
  Quantum Field Theory}},
  \href{https://doi.org/10.1007/JHEP04(2010)008}{\emph{JHEP} {\bfseries 04}
  (2010) 008} [\href{https://arxiv.org/abs/1001.4815}{{\ttfamily 1001.4815}}].

\bibitem{Wu:2010yr}
J.~Wu, J.A.~Hutasoit, D.~Boyanovsky and R.~Holman, \emph{{Neutrino
  Oscillations, Entanglement and Coherence: A Quantum Field theory Study in
  Real Time}}, \href{https://doi.org/10.1142/S0217751X11054954}{\emph{Int. J.
  Mod. Phys. A} {\bfseries 26} (2011) 5261}
  [\href{https://arxiv.org/abs/1002.2649}{{\ttfamily 1002.2649}}].

\bibitem{Akhmedov:2012uu}
E.~Akhmedov, D.~Hernandez and A.~Smirnov, \emph{{Neutrino production coherence
  and oscillation experiments}},
  \href{https://doi.org/10.1007/JHEP04(2012)052}{\emph{JHEP} {\bfseries 04}
  (2012) 052} [\href{https://arxiv.org/abs/1201.4128}{{\ttfamily 1201.4128}}].

\bibitem{Ettefaghi:2020otb}
M.M.~Ettefaghi, Z.S.~Tabatabaei~Lotfi and R.~Ramezani~Arani, \emph{{Quantum
  correlations in neutrino oscillation: Coherence and entanglement}},
  \href{https://doi.org/10.1209/0295-5075/132/31002}{\emph{EPL} {\bfseries 132}
  (2020) 31002} [\href{https://arxiv.org/abs/2011.13010}{{\ttfamily
  2011.13010}}].

\bibitem{Blasone:2021cau}
M.~Blasone, S.~De~Siena and C.~Matrella, \emph{{Wave packet approach to quantum
  correlations in neutrino oscillations}},
  \href{https://doi.org/10.1140/epjc/s10052-021-09471-4}{\emph{Eur. Phys. J. C}
  {\bfseries 81} (2021) 660}
  [\href{https://arxiv.org/abs/2104.03166}{{\ttfamily 2104.03166}}].

\bibitem{Krueger:2023skk}
R.~Krueger and T.~Schwetz, \emph{{Decoherence effects in reactor and Gallium
  neutrino oscillation experiments: a QFT approach}},
  \href{https://doi.org/10.1140/epjc/s10052-023-11711-8}{\emph{Eur. Phys. J. C}
  {\bfseries 83} (2023) 578}
  [\href{https://arxiv.org/abs/2303.15524}{{\ttfamily 2303.15524}}].

\bibitem{Giunti:1991ca}
C.~Giunti, C.W.~Kim and U.W.~Lee, \emph{{When do neutrinos really oscillate?:
  Quantum mechanics of neutrino oscillations}},
  \href{https://doi.org/10.1103/PhysRevD.44.3635}{\emph{Phys. Rev. D}
  {\bfseries 44} (1991) 3635}.

\bibitem{Giunti:1991sx}
C.~Giunti, C.W.~Kim and U.W.~Lee, \emph{{Coherence of neutrino oscillations in
  vacuum and matter in the wave packet treatment}},
  \href{https://doi.org/10.1016/0370-2693(92)90308-Q}{\emph{Phys. Lett. B}
  {\bfseries 274} (1992) 87}.

\bibitem{Giunti:1997wq}
C.~Giunti and C.W.~Kim, \emph{{Coherence of neutrino oscillations in the wave
  packet approach}},
  \href{https://doi.org/10.1103/PhysRevD.58.017301}{\emph{Phys. Rev. D}
  {\bfseries 58} (1998) 017301}
  [\href{https://arxiv.org/abs/hep-ph/9711363}{{\ttfamily hep-ph/9711363}}].

\bibitem{Giunti:2000kw}
C.~Giunti and C.W.~Kim, \emph{{Quantum mechanics of neutrino oscillations}},
  \href{https://doi.org/10.1023/A:1012230026160}{\emph{Found. Phys. Lett.}
  {\bfseries 14} (2001) 213}
  [\href{https://arxiv.org/abs/hep-ph/0011074}{{\ttfamily hep-ph/0011074}}].

\bibitem{Beuthe:2002ej}
M.~Beuthe, \emph{{Towards a unique formula for neutrino oscillations in
  vacuum}}, \href{https://doi.org/10.1103/PhysRevD.66.013003}{\emph{Phys. Rev.
  D} {\bfseries 66} (2002) 013003}
  [\href{https://arxiv.org/abs/hep-ph/0202068}{{\ttfamily hep-ph/0202068}}].

\bibitem{Kayser:2010pr}
B.~Kayser and J.~Kopp, \emph{{Testing the Wave Packet Approach to Neutrino
  Oscillations in Future Experiments}},
  \href{https://arxiv.org/abs/1005.4081}{{\ttfamily 1005.4081}}.

\bibitem{Ioannisian:1998ch}
A.~Ioannisian and A.~Pilaftsis, \emph{{Neutrino oscillations in space within a
  solvable model}},
  \href{https://doi.org/10.1103/PhysRevD.59.053003}{\emph{Phys. Rev. D}
  {\bfseries 59} (1999) 053003}
  [\href{https://arxiv.org/abs/hep-ph/9809503}{{\ttfamily hep-ph/9809503}}].

\bibitem{Karamitros:2022nnh}
D.~Karamitros and A.~Pilaftsis, \emph{{Toward a localized S-matrix theory}},
  \href{https://doi.org/10.1103/PhysRevD.108.036007}{\emph{Phys. Rev. D}
  {\bfseries 108} (2023) 036007}
  [\href{https://arxiv.org/abs/2208.10425}{{\ttfamily 2208.10425}}].

\bibitem{deGouvea:2020hfl}
A.~de~Gouvea, V.~de~Romeri and C.A.~Ternes, \emph{{Probing neutrino quantum
  decoherence at reactor experiments}},
  \href{https://doi.org/10.1007/JHEP08(2020)049}{\emph{JHEP} {\bfseries 08}
  (2020) 018} [\href{https://arxiv.org/abs/2005.03022}{{\ttfamily
  2005.03022}}].

\bibitem{deGouvea:2021uvg}
A.~de~Gouv\^ea, V.~De~Romeri and C.A.~Ternes, \emph{{Combined analysis of
  neutrino decoherence at reactor experiments}},
  \href{https://doi.org/10.1007/JHEP06(2021)042}{\emph{JHEP} {\bfseries 06}
  (2021) 042} [\href{https://arxiv.org/abs/2104.05806}{{\ttfamily
  2104.05806}}].

\bibitem{Jones:2022hme}
B.J.P.~Jones, E.~Marzec and J.~Spitz, \emph{{Width of a beta-decay-induced
  antineutrino wave packet}},
  \href{https://doi.org/10.1103/PhysRevD.107.013008}{\emph{Phys. Rev. D}
  {\bfseries 107} (2023) 013008}
  [\href{https://arxiv.org/abs/2211.00026}{{\ttfamily 2211.00026}}].

\bibitem{Smirnov:2022aab}
A.Y.~Smirnov, \emph{{Neutrino oscillations unlocked}},
  \href{https://doi.org/10.22323/1.421.0001}{\emph{PoS} {\bfseries NOW2022}
  (2023) 001} [\href{https://arxiv.org/abs/2212.10242}{{\ttfamily
  2212.10242}}].

\bibitem{Weinberg:1995mt}
S.~Weinberg, \emph{{The Quantum theory of fields. Vol. 1: Foundations}},
  Cambridge University Press (6, 2005),
  \href{https://doi.org/10.1017/CBO9781139644167}{10.1017/CBO9781139644167}.

\bibitem{Ishikawa:2018koj}
K.~Ishikawa and K.-y.~Oda, \emph{{Particle decay in Gaussian wave-packet
  formalism revisited}}, \href{https://doi.org/10.1093/ptep/pty127}{\emph{PTEP}
  {\bfseries 2018} (2018) 123B01}
  [\href{https://arxiv.org/abs/1809.04285}{{\ttfamily 1809.04285}}].

\bibitem{Ishikawa:2020hph}
K.~Ishikawa, K.~Nishiwaki and K.-y.~Oda, \emph{{Scalar scattering amplitude in
  the Gaussian wave-packet formalism}},
  \href{https://doi.org/10.1093/ptep/ptaa127}{\emph{PTEP} {\bfseries 2020}
  (2020) 103B04} [\href{https://arxiv.org/abs/2006.14159}{{\ttfamily
  2006.14159}}].

\bibitem{Ishikawa:2021bzf}
K.~Ishikawa, K.~Nishiwaki and K.-y.~Oda, \emph{{New effect in wave-packet
  scatterings of quantum fields}},
  \href{https://arxiv.org/abs/2102.12032}{{\ttfamily 2102.12032}}.

\bibitem{Oda:2021tiv}
K.-y.~Oda and J.~Wada, \emph{{A complete set of Lorentz-invariant wave packets
  and modified uncertainty relation}},
  \href{https://doi.org/10.1140/epjc/s10052-021-09482-1}{\emph{Eur. Phys. J. C}
  {\bfseries 81} (2021) 751}
  [\href{https://arxiv.org/abs/2104.01798}{{\ttfamily 2104.01798}}].

\bibitem{Oda:2023qek}
K.-y.~Oda and J.~Wada, \emph{{Lorentz-covariant spinor wave packet}},
  \href{https://arxiv.org/abs/2307.05932}{{\ttfamily 2307.05932}}.

\bibitem{OdaWada2}
K.-y.~Oda and J.~Wada, \emph{{Neutrino Mixing in Lorentz-covariant Formalism of
  Spinor Wave Packet (tmp)}}, {\emph{in progress} }.

\bibitem{Ishikawa:2005zc}
K.~Ishikawa and T.~Shimomura, \emph{{Generalized S-matrix in mixed
  representations}}, \href{https://doi.org/10.1143/PTP.114.1201}{\emph{Prog.
  Theor. Phys.} {\bfseries 114} (2006) 1201}
  [\href{https://arxiv.org/abs/hep-ph/0508303}{{\ttfamily hep-ph/0508303}}].

\end{thebibliography}\endgroup

\end{document}